\documentstyle[preprint,aps,amsfonts]{revtex}
\def\mR{{\Bbb R}}
\def\mZ{{\Bbb Z}}

\input epsf

\begin{document}

\def\Re{\mathop{\rm Re}}
\def\Im{\mathop{\rm Im}}

\tighten

\preprint{gr-qc/9505041 \qquad NSF-ITP-95-15\qquad DAMTP-R95/27}

\title{$S$-Duality at the Black Hole Threshold in Gravitational Collapse}

\author{
	Douglas M.~Eardley$^1$\footnote{Electronic address: \tt
	doug@itp.ucsb.edu\hfil}
	Eric W.~Hirschmann,$^2$\footnote{Electronic address: \tt
	ehirsch@dolphin.physics.ucsb.edu\hfil}
	and
	James H.~Horne,$^3$\footnote{Electronic address: \tt
	jhh20@damtp.cam.ac.uk\hfil}}
\address{$^1$Institute for Theoretical Physics,
	University of California,
	Santa Barbara, CA 93106-4030}
\address{$^2$Dept.~of Physics,
	University of California,
	Santa Barbara, CA 93106-9530}
\address{$^3$Dept.~of Applied Mathematics and Theoretical Physics,
	Silver St,
	Cambridge CB3 9EW, UK}
\date{\today}
\maketitle
\begin{abstract}
We study gravitational collapse of the axion/dilaton field in
classical low energy string theory, at the threshold for black hole
formation.  A new critical solution is derived that is spherically
symmetric and continuously self-similar.  The universal scaling and
echoing behavior discovered by Choptuik in gravitational collapse
appear in a somewhat different form.  In particular, echoing takes the
form of $SL(2,\mR)$ rotations ({\it
cf.}~$S$-duality).  The collapse leaves behind an outgoing pulse of
axion/dilaton radiation, with nearly but not exactly flat spacetime
within it.
\end{abstract}


\narrowtext

The striking numerical results of Choptuik~\cite{Chop} on the
spherically symmetric gravitational collapse of a real scalar field,
$\phi$, have inspired an upsurge of interest in gravitational collapse
just at the threshold for formation of black holes. Further numerical
results for vacuum relativity with axial symmetry~\cite{AE} and
perfect fluids~\cite{EC,KHA,MA} suggest that the phenomena discovered
by Choptuik are not just restricted to spherically symmetric real
scalar fields.

The thought experiment employed by Choptuik, in the context of
numerical relativity, is to ``tune'' across the critical threshold in
the space of initial conditions that separates the non-black hole
endstate from the black hole endstate, and to carefully study the
critical behavior of various quantities at this threshold~\cite{Chop}.
Two distinct kinds of scaling appear.  First, just above the black
hole threshold, the black hole mass is
$M_{\rm bh}(p) \propto (p-p^*)^\gamma$, $\gamma\approx0.37$,
where $p$ is some tuning parameter of the initial conditions.

Second, exactly at the threshold, there appears a unique critical
solution acting as an attractor for all nearby initial conditions on
threshold.  This critical solution --- which we will call a
``choptuon'' --- exhibits a striking, recurrent ``echoing'' behavior:
Asymptotically at small time-scales $t$ and length-scales $r$ near the
collapse, it repeats itself at ever-decreasing scales $t'=
e^{-n\Delta}t$, $r'= e^{-n\Delta}r$ (for $n \in \mZ_+$).  Here
$\Delta$ is either a fixed constant of the solution
($\Delta\approx\log 30$ for a real scalar field~\cite{Chop},
$\Delta\approx\log 5$ for vacuum gravity~\cite{AE}) demonstrating a
discrete self-similarity; or an arbitrary constant, demonstrating a
continuous self-similarity~\cite{EC,HE}.  The choptuon is itself a
smooth solution of the field equations to the past of the asymptotic
point $(t,r)=(0,0)$, but clearly has some kind of spacetime
singularity visible from infinity at that point; for instance, the
Riemann curvature diverges at that point.

Choptuons represent in principle a means by which effects of Planck
scale gravity, even quantum gravity or superstring theory, might be
observable in the present universe.  Therefore, the quantum
gravitational and stringy generalizations of the choptuon demand
study~\cite{ST}.

In this Letter we report on threshold behavior of gravitational
collapse, and Choptuik scaling, in classical 3+1-dimensional
low-energy effective string theory: {\it i.e.,} in general relativity
coupled to a dilaton $\Phi$ and an axion $a$.  The fields $a$ and
$\Phi$ can be combined into a single complex field $\tau \equiv
a+ie^{-2\Phi}$, and then the effective action for the model is
(omitting gauge fields\footnote{One might think that the choptuon
could carry gauge charge as well as mass, however this appears not to
be so.})
\begin{equation}
S = \int d^4 x \sqrt{-g} \left( R - {1 \over 2} { \partial_a \tau
\partial^a \bar{\tau} \over (\mathop{\rm Im}\tau)^2} \right) \; .
\label{eaction}\end{equation}
The equations of motion from this action are
\begin{mathletters}
\label{eoms}
\begin{eqnarray}
R_{ab} - {1 \over 4 (\mathop{\rm Im}\tau)^2} ( \partial_a \tau \partial_b
\bar{\tau} + \partial_a \bar{\tau} \partial_b \tau) & = & 0 \; , \\
\nabla^a \nabla_a \tau + { i \nabla^a \tau \nabla_a \tau \over
\mathop{\rm Im}\tau} = 0 \; .
\end{eqnarray}
\end{mathletters}%
Because the model~(\ref{eoms}) arises from an underlying theory of
quantum gravity, it should be an excellent starting point for
investigating quantum effects.

We assume spherical symmetry.  We also assume continuous
self-similarity, which means we work exactly at the threshold for
black hole formation, and construct a critical solution, as
in~\cite{EC,HE}.  Restricting to $a=0$ in~(\ref{eoms}) reduces to the
model studied by Choptuik, which has discrete self-similarity.
Generic initial conditions for $a$ will presumably lead at threshold
either to discrete or to continuous self-similarity; further numerical
work is needed to decide.

The model (\ref{eoms}) has an extra global symmetry not present in
general relativity, an $SL(2,\mR)$ symmetry that acts on $\Phi$
and $a$, but leaves the spacetime metric invariant; this is a classical
version of the conjectured $SL(2,\mZ)$ symmetry of heterotic
string theory called $S$-duality~\cite{SenRev}.  It acts on $\tau$ as
\begin{equation}
\tau \rightarrow {a\tau+b \over c\tau+d} \; ,
\label{sltwo}\end{equation}
where $(a,b,c,d) \in \mR$ with $ad - bc = 1$, while leaving
$g_{\mu\nu}$ invariant.

The main results of this Letter are: (1) Choptuik echoing can and does
occur in this model, taking the novel form of these $SL(2,\mR)$
rotations associated with $S$-duality; (2) A critical solution exists
that is continuously self-similar, as in~\cite{EC,HE}; (3) The global
structure of the critical spacetime is determined.

By the assumption of continuous self-similarity, there exists a
homothetic Killing vector $\xi$ that generates global scale
transformations
\begin{equation}
{\cal L}_{\xi} g_{ab} = 2 g_{ab} \; .
\label{metxi}\end{equation}
How should $\tau$ transform under $\xi$?  The naive answer (from
dimensional analysis) is that $\tau$ should be invariant.  However,
quite generally, one must allow for spacetime symmetries to mix with
internal symmetries. For instance, the Maxwell equations for the
potential $A_\mu$ are conformally invariant only in the sense that a
gauge transformation be allowed to accompany each conformal
transformation.  In our case, we must allow for $\tau$ to be scale
invariant up to an infinitesimal $SL(2,\mR)$ transformation
\begin{equation}
{\cal L}_{\xi} \tau = \alpha_0 + \alpha_1 \tau + \alpha_2 \tau^2 \; ,
\label{tauxi}\end{equation}
with $\alpha_i \in \mR$, and this will be crucial as it allows
echoing to occur.

In spherical symmetry the metric can be taken as
\begin{equation}
	ds^2 = \left(1+u(t,r)\right)\left(- b(t,r)^2dt^2 + dr^2\right)
			+ r^2d\Omega^2 \; .
\label{metric}\end{equation}
The time coordinate is chosen so that gravitational collapse on the
axis of spherical symmetry first occurs at $t=0$, and the metric is
regular for $t<0$.  This metric remains invariant in form under
transformations $t=t(t')$ of the time coordinate, and this invariance
can be used to set $b(t,0)=1$ for $t<0$ on the axis.  By regularity
(no cone singularity), $u(t,0)=0$ on the axis for $t<0$ as well.
Initial conditions, assumed smooth, may be given on the hypersurface
$t=-1$.

In these coordinates $\xi = t\partial_t + r\partial_r$.  Continuous
self-similarity (\ref{metxi}) is then implemented by defining a scale
invariant coordinate $z \equiv - r/t$.  In terms of $z$,
\begin{equation}
	b(t,r) = b(z), \quad {\rm and} \quad u(t,r) = u(z),
\label{zansatz}\end{equation}
where the real functions $b(z)$, $u(z)$ are functions solely of the
scale invariant coordinate $z$. From Eq.~(\ref{tauxi}), the $\tau$
field must satisfy
\begin{equation}
 \tau(t,r)	=  i { 1 - (-t)^{i \omega} f(z) \over 1 + (-t)^{i
\omega} f(z)} ,
\label{tauansatz}
\end{equation}
where $\omega$ is a real constant, so far arbitrary, and where the
complex function $f(z)$ depends solely on $z$.\footnote{We have
adopted a global $SL(2,\mR)$ gauge here.  In
principle, $\omega$ can also be purely imaginary or zero, but
these do not seem to give interesting solutions.}

We can now use Eqs.~(\ref{zansatz},\ref{tauansatz}) in the equations
of motion~(\ref{eoms}). After some work, $u(z)$ drops out, giving the
system of ODEs
\widetext
\begin{mathletters}
\label{fzeom}
\begin{eqnarray}
0 & = & b' + { z(b^2 - z^2) \over b (1 - |f|^2)^2} f' \bar{f}' - {
i \omega (b^2 - z^2) \over b (1 - |f|^2)^2} (f \bar{f}' - \bar{f} f')
- {\omega^2 z |f|^2 \over b (1 - |f|^2)^2} \; ,\\
0 & = & f''
     - {z (b^2 + z^2) \over b^2 (1 - |f|^2)^2} f'^2 \bar{f}'
     + {2 \over (1 - |f|^2)} \left(1
       - {i \omega (b^2 + z^2) \over 2 b^2 (1 - |f|^2)} \right) \bar{f} f'^2
     + {i \omega (b^2 + 2 z^2) \over b^2 (1 - |f|^2)^2} f f'
\bar{f}' \nonumber \\
 && + {2 \over z} \left(1 + {i \omega z^2 (1 + |f|^2) \over (b^2 - z^2)
(1 - |f|^2)} + {\omega^2 z^4 |f|^2 \over b^2 (b^2 - z^2) (1 -
|f|^2)^2}\right) f' + {\omega^2 z \over b^2 (1 - |f|^2)^2} f^2
\bar{f}' \nonumber \\
&& + {i \omega \over (b^2 - z^2)} \left(1 - {i \omega (1 + |f|^2)
\over (1 - |f|^2)} - {\omega^2 z^2 |f|^2 \over b^2 (1 - |f|^2)^2}
\right) f \; .
\end{eqnarray}
\end{mathletters}
\narrowtext

This system apparently has five singular points, as follows.  First,
the points $z=\pm0$ just represent the usual axis $r=0$ of spherical
symmetry, and regularity is easily imposed.

The point $z=\infty$ corresponds to the hypersurface $t=0$.  However,
spacetime should be smooth on this hypersurface, except at the axis,
since it lies in the Cauchy development of the initial conditions.
Indeed, the system~(\ref{fzeom}) can be rendered regular at $z=\infty$
by the following change of variables:
\begin{mathletters}
\label{new}
\begin{eqnarray}
dw  & = & b(z){dz\over z^2}, \quad w=0{\rm{\quad at}\quad}z=\infty \; ,\\
   F(w)	& = & z^{-i \omega} f(z) \; ,\\
   v(w)	& = & {b(z)\over z} \; ,			\\
   u(w) & = & u(z) \; .
\end{eqnarray}
\end{mathletters}%
In terms of the new independent variable $w$, the equations of motion
become
\widetext
\begin{mathletters}
\label{emw}
\begin{eqnarray}
0 & = & v' + {v^2 - 1 \over (1 - |F|^2)^2} F' \bar{F}' + 1 - {\omega^2
|F|^2 \over (1 - |F|^2)^2} \; ,\\
0 & = & F'' -  {2 v F'+ i \omega F \over (1 - |F|^2)^2} F' \bar{F}' +
  {2 \bar{F} F'^2 \over 1 - |F|^2}
  + {2 i \omega v \over (v^2 - 1) (1 - |F|^2)} \left( 1 + |F|^2 -
  {i \omega |F|^2 \over 1 - |F|^2} \right) F'
\nonumber \\
&&
+ {i \omega \over v^2 -1 } \left( 1 + {i \omega (1 + |F|^2)
  \over 1 - |F|^2 } + {\omega^2 |F|^2 \over (1 - |F|^2)^2}
  \right) F \; .
\end{eqnarray}
\end{mathletters}
\narrowtext

Finally, there are singular points wherever $b(z)=\pm z$ (or
$v(z)=\pm1$), at which the homothetic killing vector becomes null;
such characteristic hypersurfaces always occur in similarity solutions
to hyperbolic equations, and in this case describe light cones in
spacetime, at which new data could be injected in the $\tau$ field.
The singular point $b(z_{\scriptscriptstyle +}) = z_{\scriptscriptstyle +}$
(or $v(w_{\scriptscriptstyle +}) = 1$) describes the past
light cone of the spacetime singularity at $(t,r)=(0,0)$; see Fig.~1.
Since this cone is within the Cauchy development of the spacetime
initial conditions, the solution must be smooth across it.
\global\firstfigfalse
\begin{figure}
\centerline{\epsfysize=4in\epsfbox{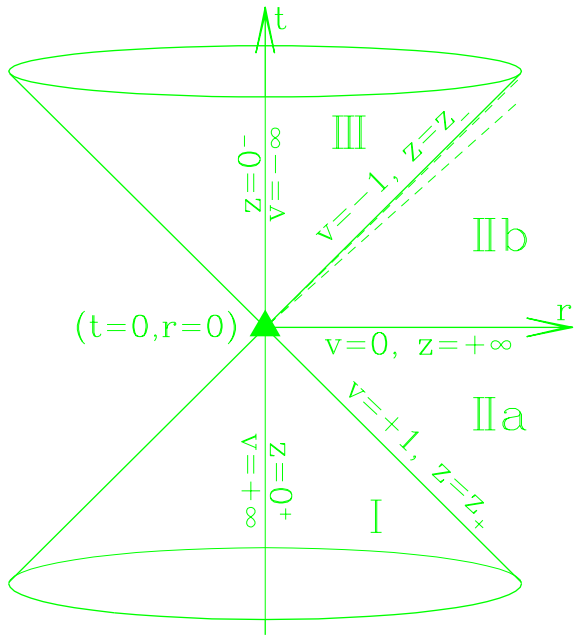}}
\caption{Interpretation of the axion/dilaton choptuon.  There are
horizons at $v=+1$ (characteristic hypersurface) and $v=-1$ (Cauchy
horizon).  There is a spacetime singularity at $(t=0,r=0)$ which is a
single point in the causal structure.  Region I is a collapsing sphere
of gravitationally bound field; in Regions IIa and IIb this blends
smoothly into an outgoing wave.  The outgoing wave oscillates
infinitely many times approaching the Cauchy horizon at $v=-1$. The
dashed lines indicate the first two oscillations.  Region III is
nearly but not precisely flat.}
\label{fig1}
\end{figure}

The singular point $b(z_{\scriptscriptstyle -}) =
-z_{\scriptscriptstyle -}$ or $v(w_{\scriptscriptstyle -}) = -1$
describes the future light cone of the spacetime singularity; see Fig.~1.
This cone is not within the Cauchy development of the spacetime initial
conditions; in particular, it is the Cauchy horizon.  Therefore the
solution may not be smooth there, and we should not impose any
boundary conditions beyond continuity of $\tau$.  Observers on this
cone will see data coming from the singularity, and indeed it is a
very interesting question to ask what they will see.  Evolution across
a Cauchy horizon can never be unique, but we will show below that
there exists a unique extension that is spherically symmetric,
self-similar, and smooth on the future axis $t>0$.  We expect that a
generic subcritical solution near the choptuon will be similar to the
extension through $z_{\scriptscriptstyle -}$, and a generic supercritical
solution will
form a singularity near $z_{\scriptscriptstyle -}$.

The main advantage of the self-similarity hypothesis is that it
reduces the equations of motion~(\ref{eoms}) to ordinary differential
equations, (\ref{fzeom}) and (\ref{emw}). Unfortunately, these ODE's
are still too complicated to solve explicitly.  Thus we use numerical
methods to determine a solution.

We are looking for a solution that is well defined everywhere, and
smooth everywhere except perhaps at the Cauchy horizon
$z_{\scriptscriptstyle -}$.  Define
two real functions $f_m(z)$ and $f_a(z)$ such that $f(z) = f_m(z) e^{i
f_a(z)}$. At any regular point $z_r$ of~(\ref{fzeom}), we need to
provide the initial conditions $b(z_r)$, $f_m(z_r)$, $f_m'(z_r)$, and
$f_a'(z_r)$ to begin numerical integration (Eq.~(\ref{fzeom}) is
independent of $f_a(z_r)$).  At any singular point, we must take care
to ensure that the fields (and their first derivatives) are
continuous.  As mentioned before, rescaling $t$ allows us to set
$b(z=0) = 1$ when $t < 0$. Requiring regularity at $z=0$ then fixes
$f_m'(0)$ and $f_a'(0)$ in terms of $f_m(0) = |f(0)|$. A Taylor
expansion of the fields $b(z)$ and $f(z)$ around $z=0$ then provides
initial conditions for integrating outwards from $z=\epsilon_0$, a
small number.

We can numerically integrate out from $z=\epsilon_0$ until the system
reaches the next singular point at $z_{\scriptscriptstyle +}$ where
$b(z_{\scriptscriptstyle +}) = z_{\scriptscriptstyle +}$.  The system
must be smooth at $z_{\scriptscriptstyle +}$, but this is numerically
difficult to impose by just integrating out from $z=\epsilon_0$. So
instead, we require that $b$ and $f$ be continuous at
$z_{\scriptscriptstyle +}$, and perform another Taylor expansion around
$z_{\scriptscriptstyle +}$, demanding smoothness.  This provides
initial conditions for integrating inwards from $z=z_{\scriptscriptstyle
+} - \epsilon_{\scriptscriptstyle +}$ in terms of
$|f(z_{\scriptscriptstyle +})|$ and $z_{\scriptscriptstyle +}$. We
then try to match the integration out from $z=\epsilon_0$ to the
integration in from $z=z_{\scriptscriptstyle +} -
\epsilon_{\scriptscriptstyle +}$ at an intermediate point, which we
choose to be $z=1$. Each integration gives four final conditions, and
there are four parameters to adjust, $|f(0)|$, $\omega$,
$z_{\scriptscriptstyle +}$, and $|f(z_{\scriptscriptstyle +})|$, so
we have a well-determined system of equations. We find that if
\begin{mathletters}
\label{zpans}
\begin{eqnarray}
\omega & = & 1.17695272200 \pm 0.00000000270 \; ,\\
z_{\scriptscriptstyle +} & = &  2.60909347510 \pm 0.00000000216 \; ,\\
|f(0)| &= & 0.892555411872 \pm 0.000000000224 \; ,\\
|f(z_{\scriptscriptstyle +})| &= & 0.364210875022 \pm 0.000000000760 \; ,
\end{eqnarray}
\end{mathletters}%
then the equations are smooth at $z=0$ and $z=z_{\scriptscriptstyle +}$.

The uncertainties quoted in~(\ref{zpans}) represent one $\sigma$
numerical uncertainties. A good understanding of the accuracy of the
parameters is crucial especially near $z_{\scriptscriptstyle -}$ where
the spacetime turns
out to be nearly, but not quite flat. To determine the errors
in~(\ref{zpans}), we have performed a large number of matchings using
a powerful numerical integrator, starting each time with different
initial estimates for the parameters, and with different values for
$\epsilon$ and the error tolerances for the integrator. All Taylor
expansions are carried out to fourth order, and should not introduce
any further uncertainties.

The above equations uniquely determine the system at $z=
z_{\scriptscriptstyle +}$. We can therefore change to the $w,v,F$
coordinate system~(\ref{new}) and integrate from
$w=w_{\scriptscriptstyle +}$ until the next singularity at
$w=w_{\scriptscriptstyle -}$ where $v(w_{\scriptscriptstyle -}) = -1$.
This takes us through $z=\infty$, $t = 0$ into the region where $t > 0$.
Near the Cauchy horizon at $w_{\scriptscriptstyle -}$, $F \rightarrow
{\rm const}$, but $F'$ oscillates rapidly. To see whether it is
possible to continue through $w_{\scriptscriptstyle -}$, or whether it
is a true singularity, we must look at the behavior of $F$ in more detail.
Near $w_{\scriptscriptstyle -}$, eq.~(\ref{emw}b) becomes
\begin{equation}
\label{fnear}
(w - w_{\scriptscriptstyle -}) F'' - c_0 F' - c_1  = 0 \; ,
\end{equation}
which has the solution
\begin{equation}
\label{fnans}
F(w) = c_2 + c_3 (w - w_{\scriptscriptstyle -})^{1 + c_0} -
		{c_1 \over c_0} (w - w_{\scriptscriptstyle -})
\end{equation}
where the $c_i$ are complex constants. Thus, $F(w)$ will be continuous
at $w_{\scriptscriptstyle -}$ if $\Re c_0 \ge -1$, and $F'(w)$ will
be continuous at $w_{\scriptscriptstyle -}$ if $\Re c_0 \ge 0$.
{}From eq.~(\ref{emw}),
\begin{equation}
\label{cans}
\Re c_0 = { \omega^2 |F(w_{\scriptscriptstyle -})|^2 \over
		(1 - |F(w_{\scriptscriptstyle -})|^2)^2 -
\omega^2 |F(w_{\scriptscriptstyle -})|^2} =
  {u(w_{\scriptscriptstyle -}) \over 1 - u(w_{\scriptscriptstyle -})} \; .
\end{equation}
Since $|F(w_{\scriptscriptstyle -})|$ is small, $\Re c_0 \ge 0$, and
both $F(w)$ and $F'(w)$ are continuous at $w_{\scriptscriptstyle -}$.
Numerically, we find that
\begin{mathletters}
\label{wmans}
\begin{eqnarray}
w_{\scriptscriptstyle -} - w_{\scriptscriptstyle +} & =
			& 2.29540547918 \pm 0.00000000192 \; ,\\
|F(w_{\scriptscriptstyle -})| & =
			& 0.007603132483 \pm 0.000000000125 \; ,\\
u(w_{\scriptscriptstyle -}) & =
			& 0.0000800854142 \pm 0.0000000000212 \; .
\end{eqnarray}
\end{mathletters}%
Since $u$ gives the mass aspect $m = {ru \over 2(1+u)}$, the Cauchy
horizon nearly, but not quite, carries data for flat
spacetime. Because $\Re c_0$ is only slightly larger than zero, the
function $F(w)$ is just barely $C^1$ at $w_{\scriptscriptstyle -}$.
$F''$ and all higher derivatives of $F$ are discontinuous at
$w_{\scriptscriptstyle -}$. This is physically acceptable as the
spacetime only depends on $F$ and $F'$.

The remaining region to consider is $0 > z > z_{\scriptscriptstyle -}$.
Since only $f$ and $f'$ are continuous at $z=z_{\scriptscriptstyle -}$,
we can not repeat the same procedure that we used for $0 < z <
z_{\scriptscriptstyle +}$. Instead, we must begin integration at
$z= - \epsilon_0$ and integrate out to $z_{\scriptscriptstyle -}$.
Since numerically we only know $|f(z_{\scriptscriptstyle -})|$, we can
try to achieve that final condition in the integration.  As before,
rescaling $t$ allows us to set $b(0_{\scriptscriptstyle -}) = 1$ and
requiring regularity at $z=0_{\scriptscriptstyle -}$ fixes $f_m'(0)$
and $f_a'(0)$ in terms of $|f(0_{\scriptscriptstyle -})|$. This is a
standard shooting problem with one free parameter and one target.
We find that
\begin{mathletters}
\label{final}
\begin{eqnarray}
z_{\scriptscriptstyle -} & = & -1.000037262799 \pm 0.000000000113 \; ,\\
|f(0_{\scriptscriptstyle -})| & = & 0.011742998497 \pm 0.000000000195 \; .
\end{eqnarray}
\end{mathletters}%
This completes the last part of the spacetime.

In conclusion, full scale numerical work on this model is highly
desirable, for several reasons.  The solution we have given is, by
assumption, continuously self-similar.  It is important to see if
discretely self-similar solutions also exist --- and if so which is
the stronger attractor.  Discretely self-similar solutions may
manifest themselves by a form of Choptuik~\cite{Chop} ``echoing'' in
which the system is invariant after a finite scale transformation by
some scale factor $\exp(-\Delta)$, up to an $SL(2,\mR)$
transformation.

\acknowledgements

This research was supported in part by the SERC under Grant GR/H57585,
and by the NSF under Grant Nos.~PHY94-07194 and PHY90-08502.  We are
grateful to the Aspen Center for Physics, where it was begun.

\end{document}